\documentclass[12pt,preprint]{aastex}

\usepackage{mathptmx}
\usepackage{courier}
\usepackage{graphicx}
\usepackage{colordvi}
\usepackage{natbib}
\usepackage{color}

\newcommand{\FeI}{\ion{Fe}{1}}

\newcommand{\aliemail}{ali@nso.edu}
\newcommand{\mandyemail}{hagenaar@lmsal.com}
\newcommand{\rolfemail}{schliche@kis.uni-freiburg.de}
\newcommand{\danielemail}{dmueller@esa.nascom.nasa.gov}

\shortauthors{Tritschler et al.}
\shorttitle{Penumbral Net-Circular Polarization}

\begin{document}

%
%

\title{The Fine-Structure of the Net-Circular Polarization in a Sunspot Penumbra}

\author{A.~Tritschler\altaffilmark{1}, D.A.N.~M\"uller\altaffilmark{2}, R.~Schlichenmaier\altaffilmark{3} and H.J.~Hagenaar\altaffilmark{4}}
\altaffiltext{1}{National Solar Observatory/Sacramento Peak\footnote{Operated by the %
       Association of Universities for Research in Astronomy, Inc. (AURA), %
       for the National Science Foundation}, 
       P.O.~Box 62, Sunspot, NM-88349, U.S.A. ; \aliemail}

\altaffiltext{2}{European Space Agency, c/o NASA 
       Goddard Space Flight Center, Greenbelt, MD, U.S.A.; \danielemail}

\altaffiltext{3}{Kiepenheuer Institut f\"ur Sonnenphysik,
       Sch\"oneckstrasse 6, Freiburg, 79104, Germany; \rolfemail}

\altaffiltext{4}{Lockheed Martin Advanced Technology Center, Org. ADBS, Bldg. 252,
          3251 Hanover Street, Palo Alto, CA 94304, U.S.A.; \mandyemail}


%
%

\begin{abstract}
We present novel evidence for a fine structure observed in the net-circular polarization (NCP) of 
a sunspot penumbra based on spectropolarimetric measurements 
utilizing the Zeeman sensitive \FeI\ 630.2\,nm line. 
For the first time we detect a filamentary organized fine structure 
of the NCP on spatial scales that are similar to the inhomogeneities found in the penumbral flow field. 
We also observe an additional property of the visible NCP, a zero-crossing of the 
NCP in the outer parts of the center-side penumbra, which has not been recognized before. 
In order to interprete the observations we solve the radiative transfer equations
for polarized light in a model penumbra with embedded magnetic flux tubes.
We demonstrate that the observed zero-crossing of the NCP  
can be explained by an increased magnetic field strength inside 
magnetic flux tubes in the outer penumbra combined with a decreased magnetic field
strength in the background field. Our results strongly support the
concept of the uncombed penumbra.
\end{abstract}


%
%

\section{Introduction}\label{sec:intro}

The asymmetry of the Stokes parameters with wavelength is a powerful diagnostic tool
to probe the relation between flows and the magnetic field in the atmosphere of sunspots. 
Line asymmetries convey information about lateral (within one resolution element) 
and the line-of-sight (LOS) gradients and discontinuities of physical parameters. 
Since the first systematic characterization and interpretation of these asymmetries 
and their spatial distribution throughout a sunspot 
by \citet[and references therein]{sanchezalmeida+lites1992} 
the complementary improvements achieved in both observational 
techniques and theoretical modeling, have advanced  
our understanding of particularly the penumbral inhomogeneities.
It is general consent that a combination of non-trivial 
magnetic field and flow topologies along the LOS are the cause of the 
observed penumbral Stokes asymmetries.
A simple and instructive way to characterize the asymmetry of Stokes V profiles
is through the net-circular polarization (NCP), defined as 
$\mathcal{N}=\int V(\lambda) d\lambda$, where the
integral is performed over one spectral line only. 
The sum of the NCP of individual lines than leads to the 
well-known broad-band circular polarization (BBCP) first noticed by 
\cite{illing+landmann+mickey1974a, illing+landmann+mickey1974b}. 
The appearance of the NCP is very distinct in the penumbra 
and shows a peculiar difference when between observations in the visible or 
in the near infrared wavelength region: in the visible at \FeI\,630.2\,nm 
the NCP maps tend to be symmetric about the disk center and sun center connection line
\citep{sanchezalmeida+lites1992, westendorp+etal2001b, mueller+etal2002}
while infrared NCP maps at \FeI\,1564.8\,nm reveal an antisymmetric distribution
\citep{schlichenmaier+collados2002, bellot+balthasar+collados2004}. 
\cite{schlichenmaier+etal2002} and \cite{mueller+etal2002} 
demonstrate that the uncombed penumbral atmosphere \citep{solanki+montavon1993}
mimicked by the \textsl{moving tube model} \citep{schlichenmaier+jahn+schmidt1998a} 
reproduces the azimuthal behavior of the NCP and, furthermore, convincingly show that
the incisive discrepancy between the visible and the near infrared can be 
understood when the effects of anomalous dispersion 
\citep{landolfi+landi1996} are included. 

So far penumbral NCP maps of the \FeI\,630.2\,nm line have been only presented 
before by \cite{sanchezalmeida+lites1992} (in form of a contour line map), 
\cite{westendorp+etal2001b}, \cite{mueller+etal2002} and \cite{mueller+etal2006}.
These maps visualize very well the symmetric behaviour of the 
NCP but lack the spatial resolution to reveal the presence of 
any potential fine structure. 

In this Letter we focus on high-spatial resolution spectropolarimetric measurements of a
sunspot to investigate inhomogeneities in the penumbral magnetic field and flow field.
Instead of utilizing inversion techniques we calculate the NCP 
and line-of-sight (LOS) velocities and elaborate on the properties of the same. 
In order to interpret the observational results we use the 
3D geometric flux tube model VTUBE \citep{mueller+etal2006} 
and perform subsequent radiative transfer calculations from which 
NCP and Doppler maps can be synthesized. 


%
%

\section{Observations and Data Reduction}\label{sec:obs+data}

The spectropolarimetric observations were obtained on 2005 June 30, 
with the Diffraction Limited Spectro-Polarimeter \citep[DLSP, ][]{sankar+etal2004}
\footnote{The DLSP was built in collaboration with the High-Altitude Observatory.} 
located at the NSO Dunn Solar Telescope (DST), Sacramento Peak, New Mexico. 
The DLSP was operated in its high-resolution mode
(0\farcs0893\,pixel$^{-1}$ along the slit) and in conjunction with the
high-order adaptive optics system \citep{rimmele2004}. 
The slit width of 12\,microns is matched to the detector image 
scale along the spectrograph slit. In spectral direction the instrument samples the
solar spectrum from 630.0\,nm to 630.4\,nm with 2.1\,pm\,pixel$^{-1}$.
A two-dimensional map was generated by moving the slit across the 
leading sunspot (positive polarity) of AR NOAA 10781 located at $\mu=0.74$ (corresponding to a viewing
angle of 42$^{\circ}$) with a step size of 0\farcs089 covering 
a total field-of-view of 58\farcs9 (660 steps). 
The slit was oriented along solar North-South and the scan direction of the
DLSP was solar West-East. At each slit position 16 images of 30\,ms 
exposure time each were accumulated
for each of the four individual modulation states leading to a total integration time
of 2\,sec. A full map is accomplished within 56\,min.
The measurements are compensated for polarization effects caused by the telescope and the
polarimeter following the calibration procedure developed by \cite{skumanich+etal1997}.
The very good seeing in conjunction with the 
high-order AO system led to exceptional stable observing conditions 
while scanning the solar image. We estimate our effective spatial resolution 
in the DLSP map to be about 0\farcs4. 

We calculate maps of the net circular polarization by integrating the Stokes $V$ signal 
over the \FeI\,630.25\,nm line. 
Because of its relatively minor contamination 
by one of the two telluric lines the \FeI\,630.15\,nm line is used to derive
line-of-sight (LOS) Doppler velocities. We apply a Fourier phase method 
\citep{schmidt+stix+woehl1999} to the $I+V$ and $I-V$ profile 
of the \FeI\,630.15\,nm line to calculate line shifts. 
An absolute velocity calibration is performed by taking advantage of the telluric O$_2$ lines 
\citep[see e.g. ][]{martinezpillet+lites+skumanich1997} present in the recorded spectrum. 
In contrast to the astrophysical convention we assign a negative 
sign to redshifts and a positive sign to blueshifts: dark areas in the Dopplergram 
move away from the observer, while bright areas move towards the observer.

%
%
%

\section{Results}\label{sec:results}

\begin{figure*}[t]
  \epsscale{0.315}
  \plotone{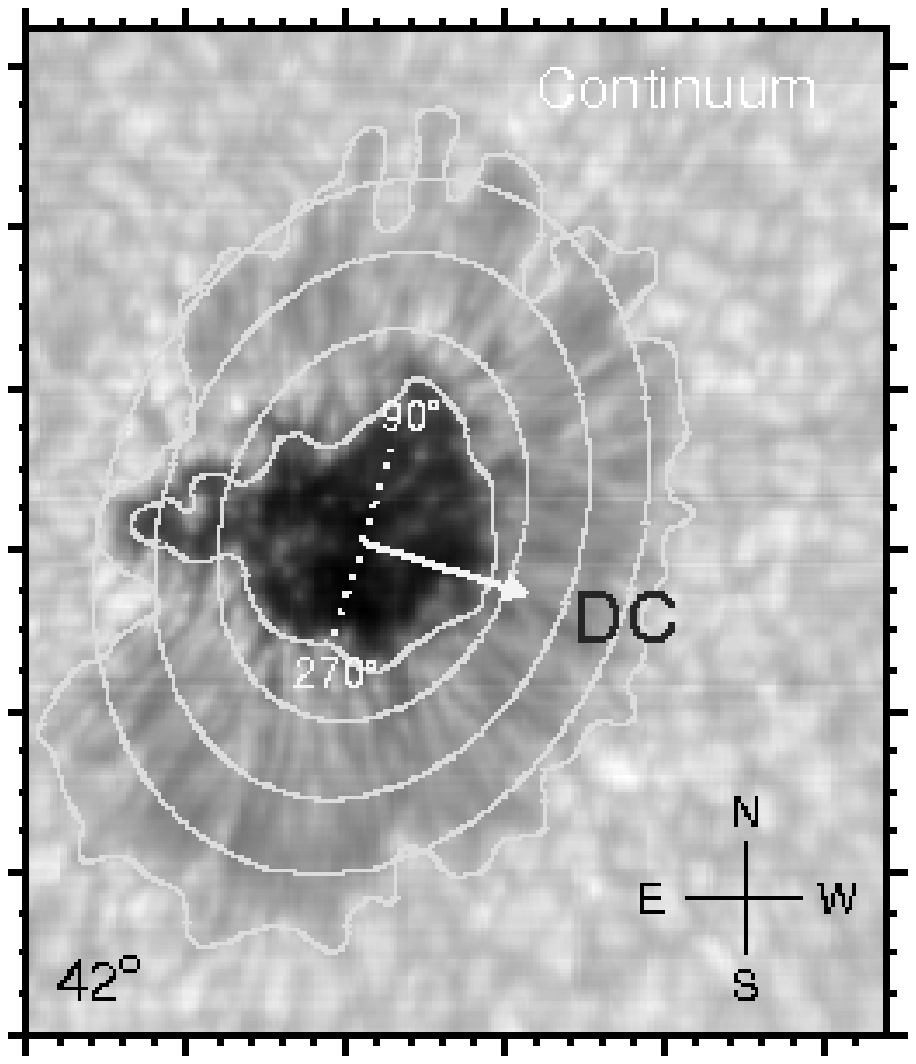}\plotone{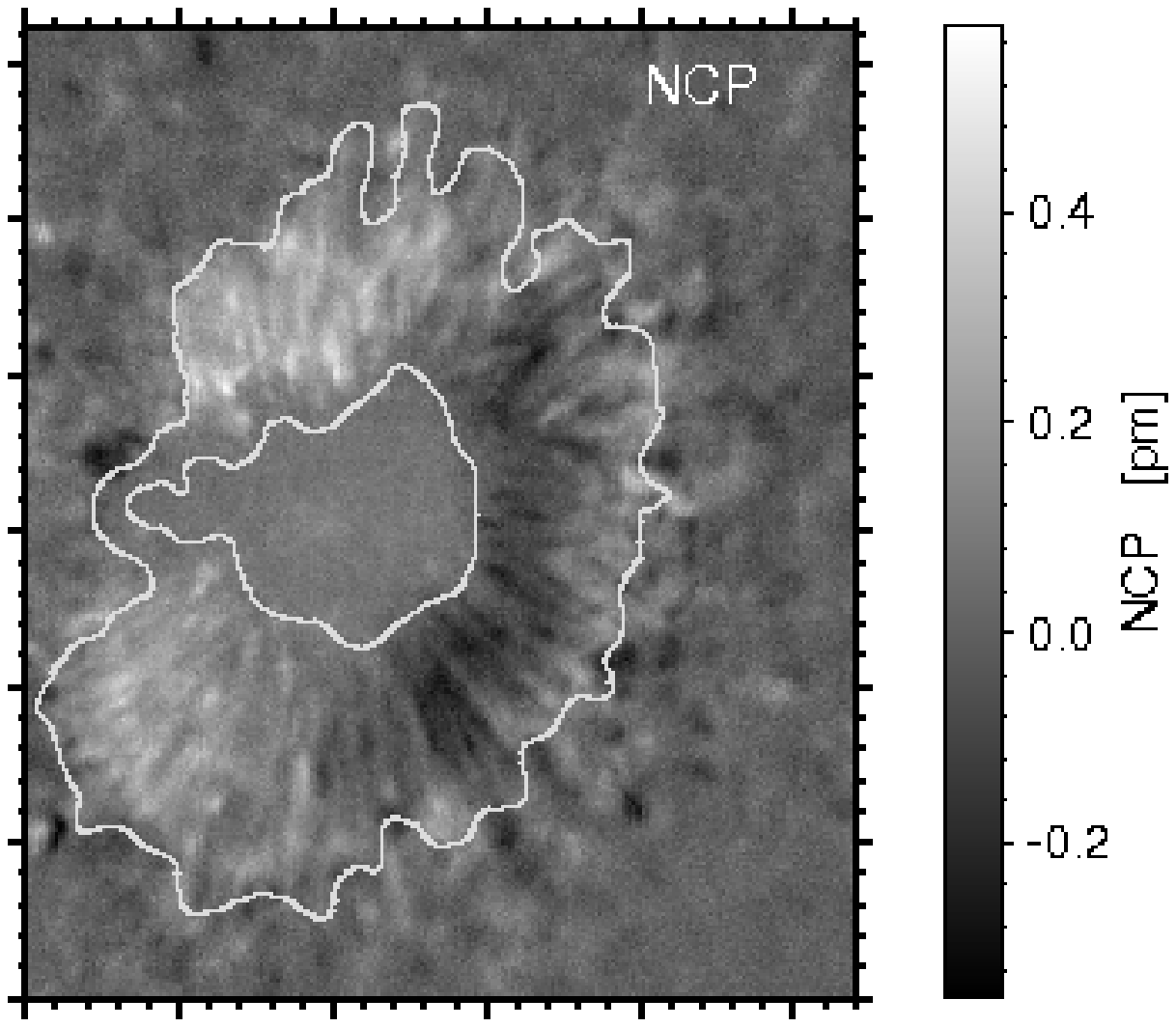}\plotone{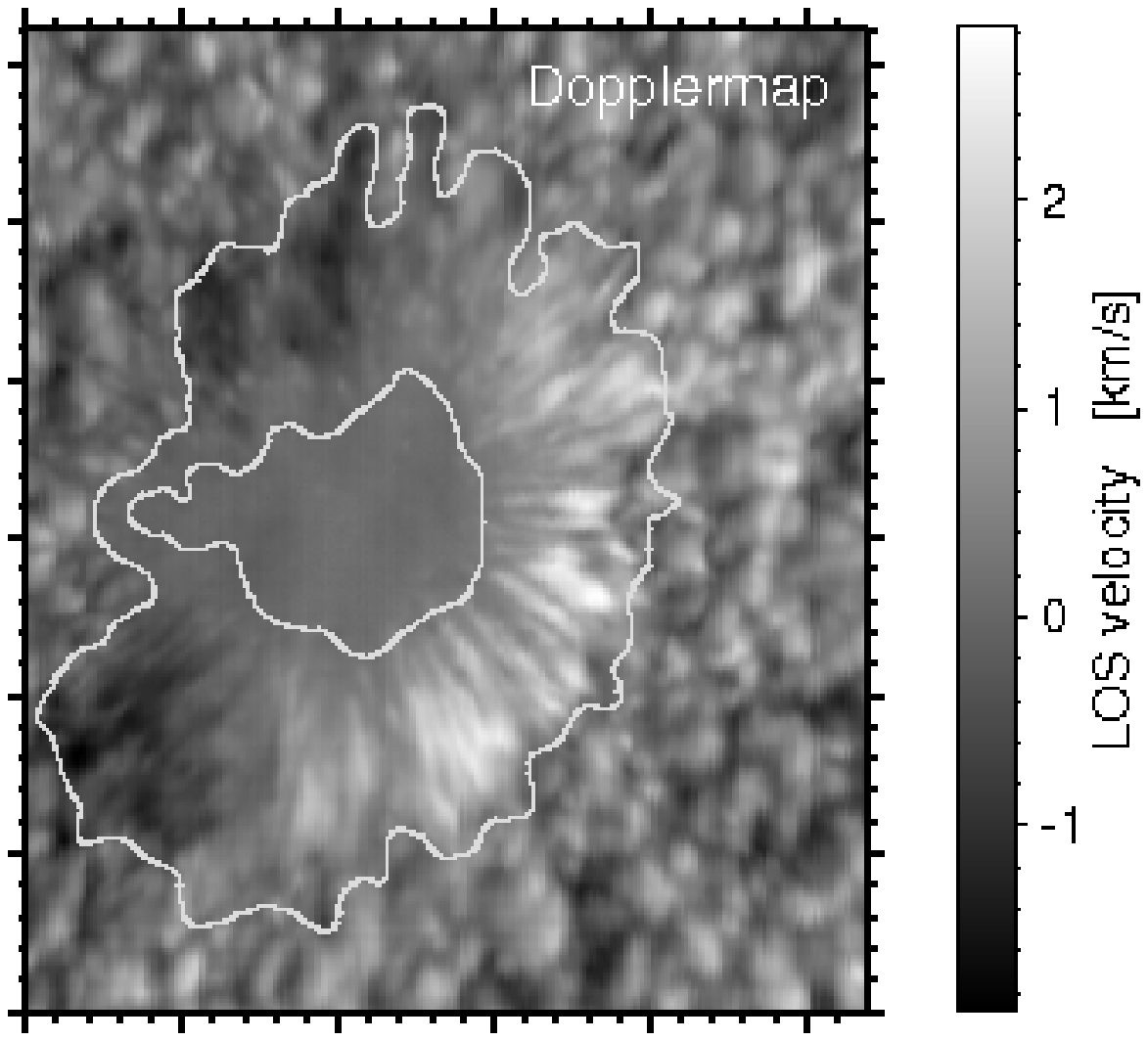}
  \caption{Continuum intensity (\textsl{left}), net circular polarization 
           (\textsl{middle}) and LOS velocity (\textsl{right}) of AR NOAA 10781 
           observed on June 30, 2005, between 14:24~UT and 15:03~UT with the DLSP. 
           The Dopplergram is calculated from the
           \FeI\,630.15\,nm line while the NCP is 
           determined from the \FeI\,630.25\,nm line.
           The arrow points towards disk center (DC). 
           The three ellipses indicate azimuthal cuts corresponding to radial distances
           from the center of the umbra of $0.5R$, $0.7R$, and $0.9R$ 
           in units of the spot radius, $R$. 
           The contour lines mark the inner and outer boundary of the
           penumbra as determined from thresholding a smoothed version of the continuum map 
           at $0.70I_c$ and $0.96I_c$ in units of the continuum intensity 
           $I_c$ of the quiet sun.}
  \label{fig:physmaps}
  \epsscale{1.0}
\end{figure*}

Well-reported properties of the penumbral NCP  
\citep[see][]{sanchezalmeida+lites1992, solanki+montavon1993}
are the sign change along azimuthal cuts centered around the spot center
and an increased magnitude on the limb side. 
Based on spectropolarimetric data with much higher spatial 
resolution than utilized before (Figure~\ref{fig:physmaps}) 
we additionally find (a) evidence for a fine-structure 
in the NCP and (b) a sign-reversal of the NCP as a function of radial 
distance from the spot center on the center-side penumbra which has not been recognized before.

Confirming previous results we find that the NCP is symmetric in the
inner and middle penumbra w.r.t.\ the line that connects the disk center with the
spot center. The same property applies to the LOS velocity map 
which indicates an intricate coupling between penumbral 
flows and the sign of the NCP. This is corroborated by the 
fine structure we detect in the NCP. 

\begin{figure*}[th]
  \epsscale{1.0}
  \plotone{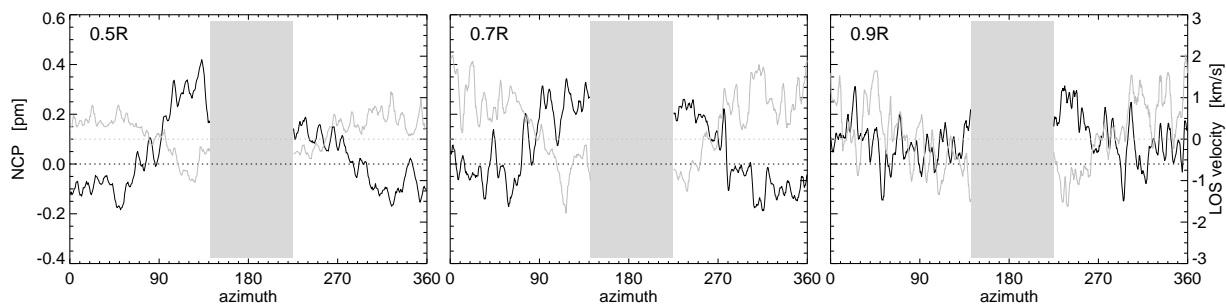}
  \caption{Azimuthal variation of the NCP (\textsl{black}) and LOS velocity (\textsl{grey}) 
           measured counterclockwise from disk center 
           direction for three different radial distances from spot center as 
           indicated in Figure~\ref{fig:physmaps} (\textsl{left}). The NCP is calculated from 
           the \FeI\,630.25\,nm line while the LOS velocity is derived from the 
           \FeI\,630.15\,nm line. The shaded areas where no data points are plotted 
           (141.5$^{\circ}$-225.0$^{\circ}$) mark the angle segment where the penumbra is missing. 
           Horizontal \textsl{dotted} lines indicate zero NCP and LOS velocity.}
  \label{fig:azimuthal}
  \epsscale{1.0}
\end{figure*}

The spatial distribution of the NCP reveals filamentary organized 
inhomogeneities throughout the penumbra with very localized enhancements 
of elongated shape on a spatial scale of 0\farcs3.
Although we detect these enhancements with no preferred location inside the 
penumbra those with the largest value of the NCP appear on the 
limb-side penumbra in proximity to where the magnetogram 
(not shown) indicates an apparent neutral line. 
They are related to anomalous (multi-lobe) Stokes V profiles 
\citep[see also ][]{sanchezalmeida+lites1992} and thus do not necessarily show up in 
the flow signal (Figure~\ref{fig:physmaps}, \textsl{right}). 
Anomalously shaped Stokes V profiles indicate either spatially unresolved contributions to the 
resolution element or LOS effects of at least two differently 
inclined magnetic components with opposite polarity that are 
shifted w.r.t.\ each other and viewed upon under an oblique angle. 
The NCP on the center-side penumbra is correlated with 
normal 2-lobe antisymmetric Stokes V profiles. On smaller spatial scales 
related to the flow channels (Figure~\ref{fig:physmaps}, \textsl{right}) 
that carry most of the Evershed flow we are unable to 
state conclusively that there is a good overall correlation 
between the NCP and the flow signatures. 
Figure~\ref{fig:azimuthal} exemplifies this statement along three azimuthal cuts
through the penumbra representative for three different radial 
distances from spot center (as indicated by the ellipses 
in Figure~\ref{fig:physmaps}, \textsl{left}). 

\begin{figure}[th]
  \epsscale{1.0}
  \plotone{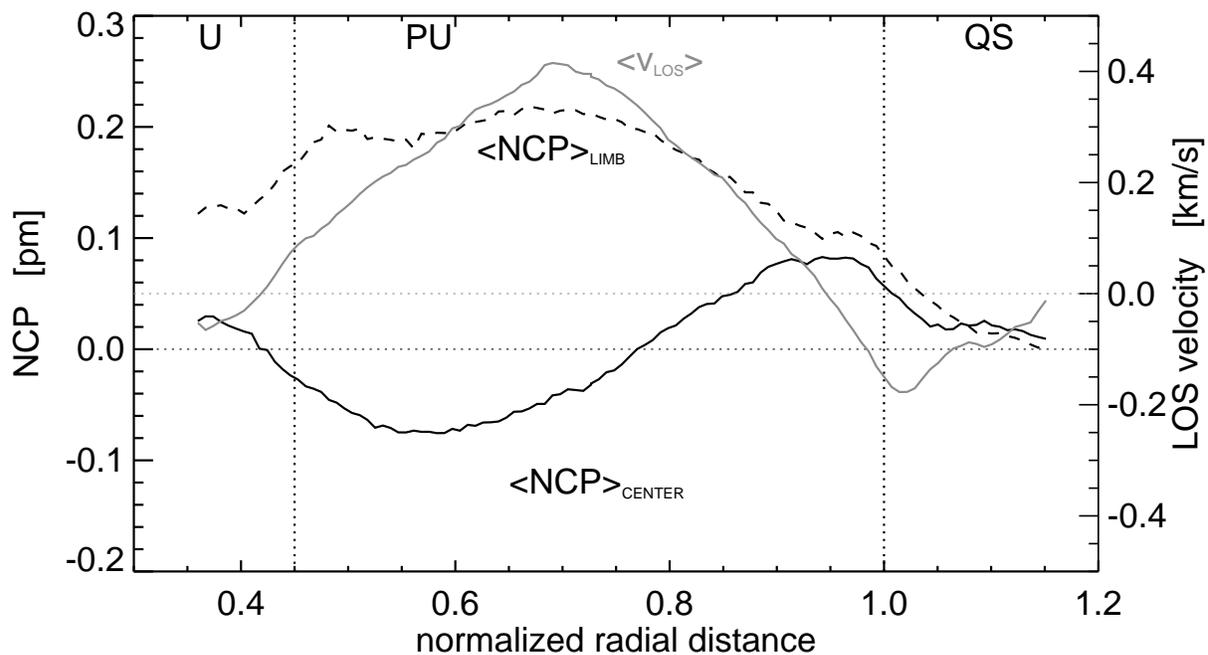}
  \caption{Radial dependence of the mean LOS velocity ($<v_{\rm LOS}>$) and 
           the NCP averaged over the limb-side ($<.>_{\rm LIMB}$) and 
           center-side ($<.>_{\rm CENTER}$), respectively. The penumbra 
           missing segment (141.5$^{\circ}$-225.0$^{\circ}$) is 
           not contributing in the averaging process. The vertical \textsl{dotted}
           lines mark the inner and outer penumbral boundary. 
           Horizontal \textsl{dotted} lines indicate zero NCP and LOS velocity. 
           The radial distance is given with respect to the spot radius.}
  \label{fig:radial}
  \epsscale{1.0}
\end{figure}

Furthermore, we detect for the first time a difference in the symmetry properties of the NCP 
observed between the outer limb- and center-side penumbra, namely a sign reversal of the
NCP in the outer center-side penumbra. \footnote{Beyond the penumbra crossing 
the visible boundary of the sunspot the NCP shows a motley behavior which we attribute 
to the presence of moving magnetic features that are known 
to show asymmetric Stokes profiles.} We verify the visual impression
of the NCP zero-crossing on the center-side penumbra in Figure~\ref{fig:radial} 
by means of the radial dependence of the NCP when averaged over the 
limb-side (90$^{\circ}$-270$^{\circ}$) and center-side 
penumbra (270$^{\circ}$-90$^{\circ}$) separately. The limb-side segment 
without penumbra does not contribute to the averaging. 

A comparison of the two curves clearly demonstrates the differences between
the NCP on the limb- and center-side penumbra in radial direction. On the limb side
the NCP has one sign only throughout the penumbra while on 
the center side the NCP reverses its sign at a radial distance of $\sim$0.75$R$, 
(where $R$ denotes the sunspots radius) peaks at 0.95$R$ and then falls 
off outside the visible penumbra. It should be noted that the course 
of the radial variation of the NCP and the exact location of
the zero-crossing point within the penumbra depends also on 
the calibration for instrumental polarization, i.e. predominantly the crosstalk 
from Stokes $I$ to $V$. We estimate that our calibration for this particular
crosstalk is better than 0.1\,\%. The NCP averaged over areas where we see
no significant Stokes $V$ signal is in the range of 1.5$\times$10$^{-2}$\,pm.
Significant changes of the radial variation of the NCP, however, take place
only if the residual crosstalk from $I$ to $V$ is in the range of 1\,\% and larger. 

%
%

\section{Model Calculations}\label{sec:model}

In order to understand the observed sign reversal of the NCP, we have
carried out radiative transfer calculations from which synthetic maps
of observable quantities are obtained.  At first sight, one may
conjecture that the observed radial sign reversal of the NCP is caused
by downflows in the outer penumbra. While this is indeed a possible
solution that is compatible with the observed NCP maps, the Doppler
maps constrain the details of the model atmosphere even further.

\begin{figure*}[th]
\epsscale{0.3}
  \plotone{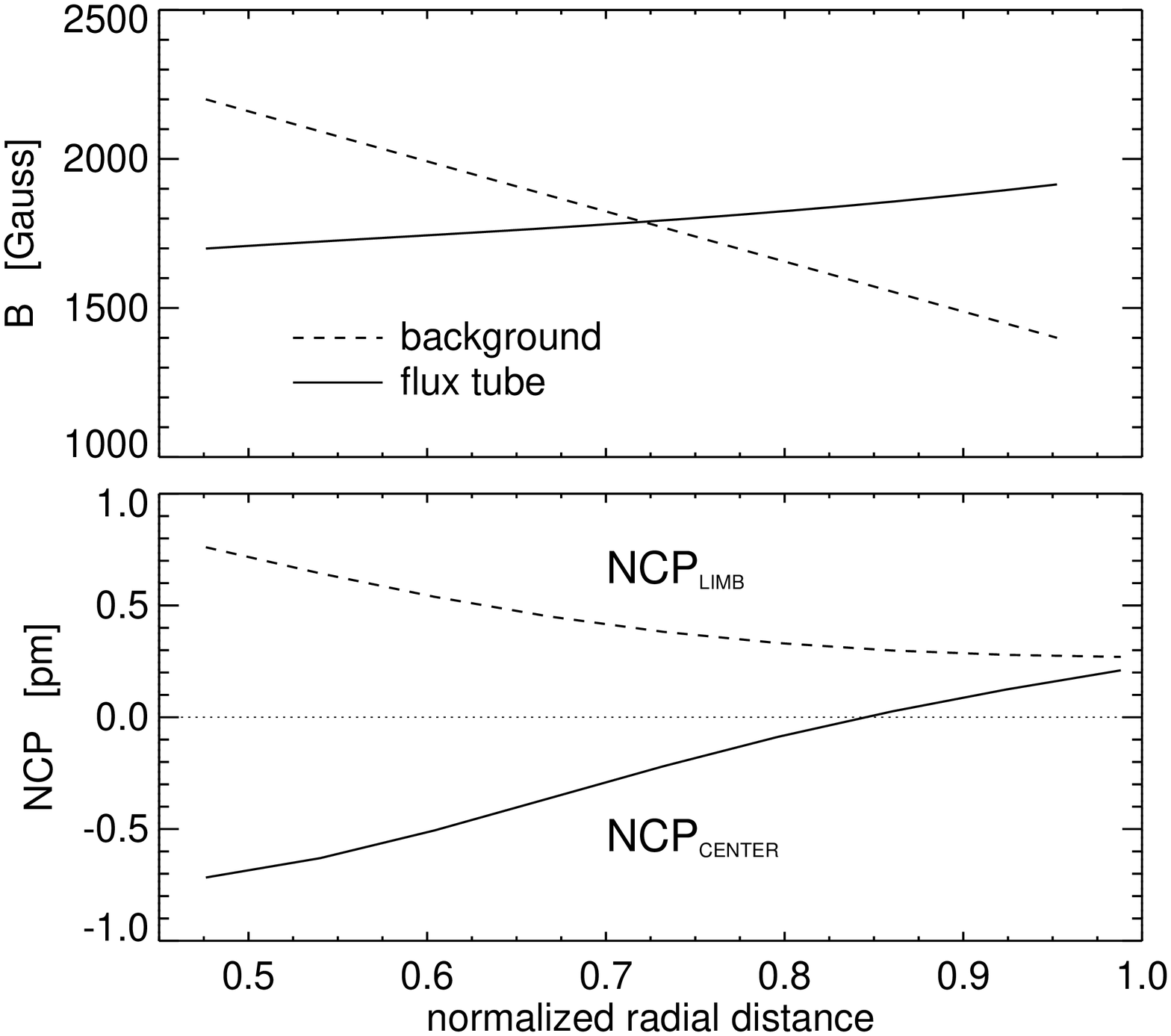}\plotone{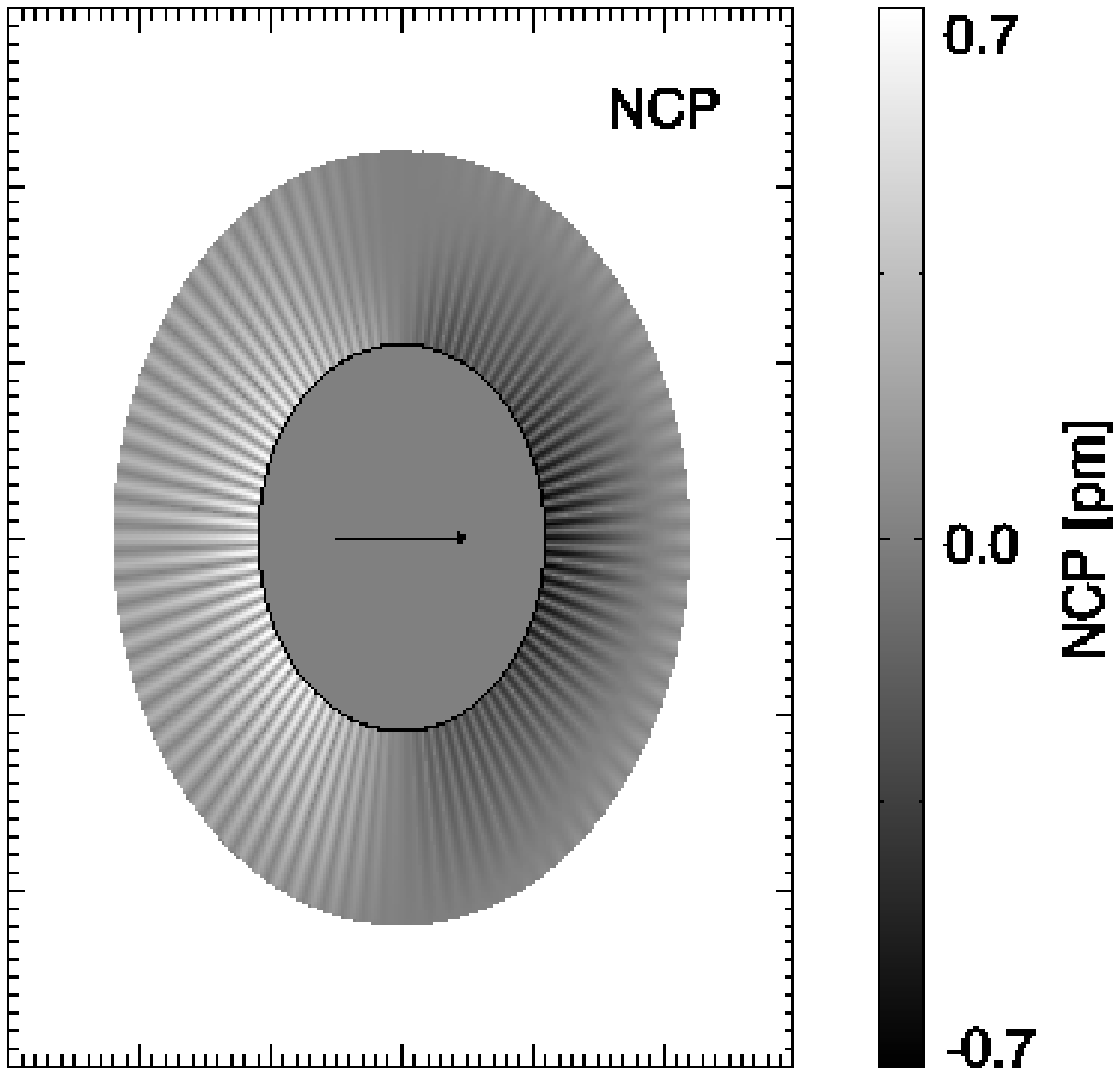}\plotone{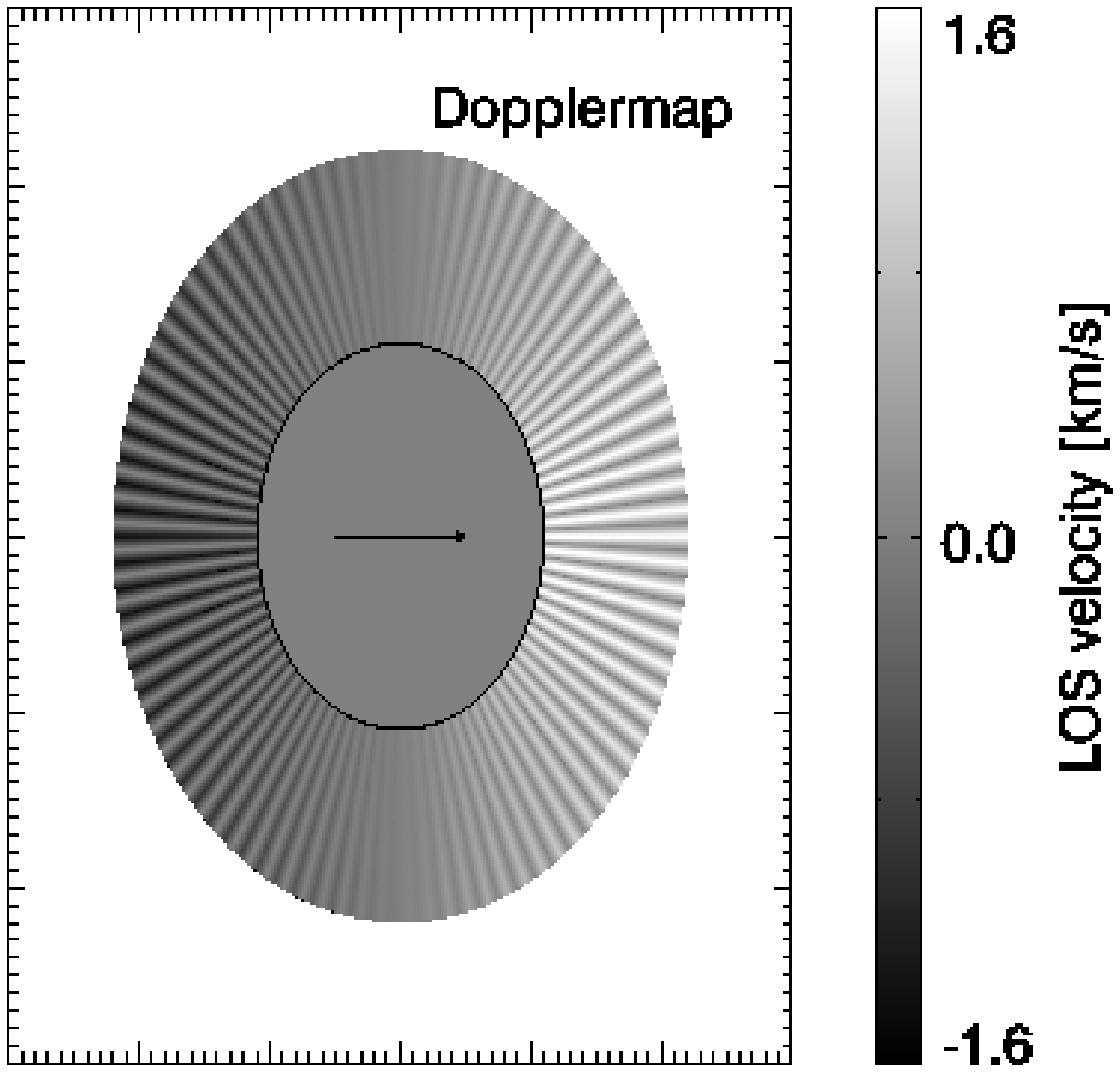}
  \caption{\textsl{Left}: Radial variation of the magnetic field strength inside the flux tube 
           and in the background field (\textsl{top}) and variation of the NCP 
           on the center- and limb-side penumbra (\textsl{bottom}). 
           Synthetic maps of the net circular polarization (\textsl {middle}) 
           and LOS velocity (\textsl{right}) of the \FeI\,630.25\,nm line 
           for a  sunspot at $\theta = 42^\circ$. To illustrate the filamentary 
           structure of the  penumbra, we alternate between flux tubes 
           with a radial width of 2$^ \circ$ and "gaps" of stationary 
           background atmosphere of the same  width.}
  \label{fig:synthetic}
  \epsscale{1.0}
\end{figure*}

It turns out that a model atmosphere with strongly downwards inclined flux
tubes in the outer penumbra is not able to explain simultaneously both the observed
sign reversal of the NCP and the constant sign of the Doppler maps on
the spot's center side. However, we find that both features can be reproduced by magnetic flux
tubes that have a magnetic field strength that is stronger than the
background magnetic field in the outer penumbra while it is weaker in the inner part 
as visualized in Figure~\ref{fig:synthetic} (\textsl{top left}).
The assumption is not unrealistic as two-component inversions of sunspots 
indicate a higher field strength in 
the tube component in the extreme outer penumbra 
\citep[e.g.][]{bellot+balthasar+collados2004}.
%

Figure~\ref{fig:synthetic} shows the synthetic 
NCP (\textsl{middle}) and Doppler (\textsl{right})
maps for a penumbral atmosphere, calculated with the VTUBE model of
\cite{mueller+etal2006}. We adopt a very generic atmospheric
configuration with horizontal magnetic flux tubes of 50\, km radius
with a flow velocity of 10\,km/s, embedded in a radially symmetric
background field as in \cite{mueller+etal2006}. 
We assume a magnetic field strength inside the flux tubes that
increases linearly from the inner penumbra (where it is 500\,G weaker
than the background field) to the outer penumbra (where it is 500\,G
stronger than the background field). It turns out that this
configuration is quite robust, and the details of the magnetic field
strength affect primarily the radial location of the zero-crossing of
the NCP. The axes of the flux tubes are placed at 30\,km above
$\tau_{500} = 1$. Raising the flux tubes mostly increases the absolute
value of the NCP and would require a lower filling factor to explain
the observed NCP values.

The effect of a jump in the magnetic field strength along the LOS on
the NCP has been studied for a simpler atmospheric configuration by
\cite{landolfi+landi1996} who derived a sign rule for the NCP due to
this so-called $\Delta B$ effect. This expression is proportional to
the difference in the magnetic field strengths between the static
background and the part of the atmosphere containing the flow. As a consequence
the NCP changes sign when $\Delta B$ changes sign.
When comparing this rule with both observational data and more realistic models,
one should note that the $\Delta B$ effect is but one of several NCP
generating effects at work. Therefore, the locations of the sign
reversals of $\Delta B$ and the NCP do not necessarily coincide. 
In particular the sum of the $\Delta B$ effect and changes in the inclination
($\Delta\gamma$ effect) results in the different behaviour of the NCP 
on the limb and center side: on the limbside
the NCP decreases with radial distance because of the increasing 
$\Delta B$ effect but does not change the sign because 
the $\Delta\gamma$ effect contributes positive to the NCP.
One should
also bear in mind the assumptions on which the findings of 
\cite{landolfi+landi1996} are based. A discussion of the results 
of \cite{landolfi+landi1996} in the context of flux tubes in the penumbra is
given by \cite{mueller2001} and \cite{mueller+etal2002}.

Our model solution is in qualitative agreement with both the observed
NCP and the observed LOS velocity. It should be pointed out though that while
the simultaneous agreement between the NCP and Doppler maps places a strong
constraint on the possible atmospheric models our solution may not 
necessarily be unique.

%
%

\section{Discussion and Conclusions}\label{sec:conclusions}

In this paper, we have presented novel evidence for a spatial fine
structure in the visible NCP of the \FeI\,630.25\,nm line. 
We find for the first time that the NCP is structured in radial filaments 
very much alike maps of other line parameters, e.g. the intensity, equivalent width 
and line width \citep[see e.g.][]{johannesson1993, 
rimmele1995a, rimmele1995b, tritschler+etal2004}. 
The spatial scales of the NCP filaments are similar to those found in the 
observed LOS velocity map although we cannot find a good overall 
correlation between the signals of the NCP and the LOS velocity. 
We conjecture that higher spatial resolution is needed to 
settle this issue. \cite{martinezpillet1997} find a correlation between the azimuthal variation of the
NCP and the magnetic field inclination on angular scales of $10^{\circ}$. Although
we cannot provide information about the inclination at this point 
we find that our observations show azimuthal variations of the NCP 
that are only of the order of $4^{\circ}$.

Furthermore, we have observed a zero-crossing of the NCP 
on the outer center-side penumbra and have 
demonstrated that this signature can be explained by an
increased magnetic field strength inside magnetic flux tubes in the
outer penumbra, while the magnetic field strength is decreased in the background
field ($\Delta B$-effect). To this end, we have compared high-resolution
observations with maps of the NCP and the LOS velocity that have been
synthesized from a 3D geometric sunspot model that has been coupled
with a 1D radiative transfer code \citep[see][ for details]{mueller+etal2006}.  
We point out that the space of possible atmospheric
models that can  reproduce the observations is significantly
constrained by  taking  into account the information contained in
Doppler maps in addition to  the NCP maps. This paper extends the work
by \cite{solanki+montavon1993}, \cite{martinezpillet2000}, \cite{mueller+etal2002}, and
\cite{mueller+etal2006} and clearly demonstrates the
need to model the "uncombed"  penumbra of sunspots in greater
detail. This would include also taking into account the effect of field line
curvature around flux tubes as considered by \cite{borrero+bellot+mueller2007}.


%
%


%
%

\clearpage

%
%
%
%
%
%
%


\begin{thebibliography}{26}
\expandafter\ifx\csname natexlab\endcsname\relax\def\natexlab#1{#1}\fi

\bibitem[{{Bellot Rubio} {et~al.}(2004){Bellot Rubio}, {Balthasar}, \&
  {Collados}}]{bellot+balthasar+collados2004}
{Bellot Rubio}, L.~R., {Balthasar}, H., \& {Collados}, M. 2004, \aap, 427, 319

\bibitem[{{Borrero} {et~al.}(2007){Borrero}, {Bellot Rubio}, \&
  {Mueller}}]{borrero+bellot+mueller2007}
{Borrero}, J.~M., {Bellot Rubio}, L.~R., \& {Mueller}, D.~A.~N. 2007, ArXiv
  e-prints, 707

\bibitem[{{Illing} {et~al.}(1974{\natexlab{a}}){Illing}, {Landman}, \&
  {Mickey}}]{illing+landmann+mickey1974a}
{Illing}, R.~M.~E., {Landman}, D.~A., \& {Mickey}, D.~L. 1974{\natexlab{a}},
  \aap, 35, 327

\bibitem[{{Illing} {et~al.}(1974{\natexlab{b}}){Illing}, {Landman}, \&
  {Mickey}}]{illing+landmann+mickey1974b}
---. 1974{\natexlab{b}}, \aap, 37, 97

\bibitem[{{Johannesson}(1993)}]{johannesson1993}
{Johannesson}, A. 1993, \aap, 273, 633

\bibitem[{{Landolfi} \& {Landi degl'Innocenti}(1996)}]{landolfi+landi1996}
{Landolfi}, M., \& {Landi degl'Innocenti}, E. 1996, \solphys, 164, 191

\bibitem[{{Martinez Pillet}(1997)}]{martinezpillet1997}
{Martinez Pillet}, V. 1997, in ASP Conf. Ser. 118,
 Advances in Physics of Sunspots, ed. B.~{Schmieder},
  J.~C. {del Toro Iniesta}, \& M.~{Vazquez}, 212

\bibitem[{{Mart{\'\i}nez Pillet}(2000)}]{martinezpillet2000}
{Mart{\'\i}nez Pillet}, V. 2000, A\&A, 361, 734

\bibitem[{{Martinez Pillet} {et~al.}(1997){Martinez Pillet}, {Lites}, \&
  {Skumanich}}]{martinezpillet+lites+skumanich1997}
{Martinez Pillet}, V., {Lites}, B.~W., \& {Skumanich}, A. 1997, \apj, 474, 810

\bibitem[{{M\"uller}(2001)}]{mueller2001}
{M\"uller}, D. 2001, Polarisation von Linien im Spektrum einer
  Sonnenfleckenpenumbra. Diploma Thesis, University of Freiburg

\bibitem[{{M{\"u}ller} {et~al.}(2006){M{\"u}ller}, {Schlichenmaier}, {Fritz},
  \& {Beck}}]{mueller+etal2006}
{M{\"u}ller}, D.~A.~N., {Schlichenmaier}, R., {Fritz}, G., \& {Beck}, C. 2006,
  \aap, 460, 925

\bibitem[{{M{\"u}ller} {et~al.}(2002){M{\"u}ller}, {Schlichenmaier}, {Steiner},
  \& {Stix}}]{mueller+etal2002}
{M{\"u}ller}, D.~A.~N., {Schlichenmaier}, R., {Steiner}, O., \& {Stix}, M.
  2002, \aap, 393, 305

\bibitem[{{Rimmele}(1995{\natexlab{a}})}]{rimmele1995a}
{Rimmele}, T.~R. 1995{\natexlab{a}}, \aap, 298, 260

\bibitem[{{Rimmele}(1995{\natexlab{b}})}]{rimmele1995b}
---. 1995{\natexlab{b}}, \apj, 445, 511

\bibitem[{{Rimmele}(2004)}]{rimmele2004}
---. 2004, \apj, 604, 906

\bibitem[{{S\'anchez Almeida} \& {Lites}(1992)}]{sanchezalmeida+lites1992}
{S\'anchez Almeida}, J., \& {Lites}, B.~W. 1992, ApJ, 398, 359

\bibitem[{{Sankarasubramanian} {et~al.}(2004){Sankarasubramanian}, {Gullixson},
  {Hegwer}, {Rimmele}, {Gregory}, {Spence}, {Fletcher}, {Richards}, {Rousset},
  {Lites}, {Elmore}, {Streander}, \& {Sigwarth}}]{sankar+etal2004}
{Sankarasubramanian}, K., {Gullixson}, C., {Hegwer}, S., {Rimmele}, T.~R.,
  {Gregory}, S., {Spence}, T., et al. 2004, Proceedings SPIE, 5171, 207-218

\bibitem[{{Schlichenmaier} \& {Collados}(2002)}]{schlichenmaier+collados2002}
{Schlichenmaier}, R., \& {Collados}, M. 2002, \aap, 381, 668

\bibitem[{{Schlichenmaier} {et~al.}(1998){Schlichenmaier}, {Jahn}, \&
  {Schmidt}}]{schlichenmaier+jahn+schmidt1998a}
{Schlichenmaier}, R., {Jahn}, K., \& {Schmidt}, H.~U. 1998, \apjl, 493, L121

\bibitem[{{Schlichenmaier} {et~al.}(2002){Schlichenmaier}, {M{\"u}ller},
  {Steiner}, \& {Stix}}]{schlichenmaier+etal2002}
{Schlichenmaier}, R., {M{\"u}ller}, D.~A.~N., {Steiner}, O., \& {Stix}, M.
  2002, \aap, 381, L77

\bibitem[{{Schmidt} {et~al.}(1999){Schmidt}, {Stix}, \&
  {W{\"o}hl}}]{schmidt+stix+woehl1999}
{Schmidt}, W., {Stix}, M., \& {W{\"o}hl}, H. 1999, \aap, 346, 633

\bibitem[{{Skumanich} {et~al.}(1997){Skumanich}, {Lites}, {Martinez Pillet}, \&
  {Seagraves}}]{skumanich+etal1997}
{Skumanich}, A., {Lites}, B.~W., {Martinez Pillet}, V., \& {Seagraves}, P.
  1997, \apjs, 110, 357

\bibitem[{{Solanki} \& {Montavon}(1993)}]{solanki+montavon1993}
{Solanki}, S.~K., \& {Montavon}, C. A.~P. 1993, \aap, 275, 283

\bibitem[{{Tritschler} {et~al.}(2004){Tritschler}, {Schlichenmaier}, {Bellot
  Rubio}, {the KAOS Team}, {Berkefeld}, \& {Schelenz}}]{tritschler+etal2004}
{Tritschler}, A., {Schlichenmaier}, R., {Bellot Rubio}, L.~R., {the KAOS Team},
  {Berkefeld}, T., \& {Schelenz}, T. 2004, \aap, 415, 717

\bibitem[{{Westendorp Plaza} {et~al.}(2001){Westendorp Plaza}, {del Toro
  Iniesta}, {Ruiz Cobo}, {Mart{\'\i}nez Pillet}, {Lites}, \&
  {Skumanich}}]{westendorp+etal2001b}
{Westendorp Plaza}, C., {del Toro Iniesta}, J.~C., {Ruiz Cobo}, B.,
  {Mart{\'\i}nez Pillet}, V., {Lites}, B.~W., \& {Skumanich}, A. 2001, \apj,
  547, 1130

\end{thebibliography}
\end{document}